\pdfoutput=1
\documentclass[sigconf]{acmart}
\AtBeginDocument{%
  \providecommand\BibTeX{{%
    \normalfont B\kern-0.5em{\scshape i\kern-0.25em b}\kern-0.8em\TeX}}}

\copyrightyear{2022}
\acmYear{2022}
\setcopyright{acmcopyright}
\acmConference[ICSE-SEIP '22]{44nd International Conference on Software Engineering: Software Engineering in Practice}{May 21--29, 2022}{Pittsburgh, PA, USA}
\acmBooktitle{44nd International Conference on Software Engineering: Software Engineering in Practice (ICSE-SEIP '22), May 21--29, 2022, Pittsburgh, PA, USA}
\acmPrice{15.00}
\acmDOI{10.1145/3510457.3513030}
\acmISBN{978-1-4503-9226-6/22/05}

\usepackage{subcaption}
\usepackage{multirow}
\usepackage{enumitem}
\usepackage{amsmath,amsfonts}
\usepackage{algorithmic}
\usepackage{graphicx}
\usepackage{wrapfig}

\newcommand*{\ditto}{--\texttt{"}--}
\begin{document}

\title{Mining Root Cause Knowledge from Cloud Service Incident Investigations for AIOps}


\author{Amrita Saha}
\email{amrita.saha@salesforce.com}
\affiliation{%
 \institution{Salesforce Research Asia}
}

\author{Steven C.H. Hoi}
\email{shoi@salesforce.com}
\affiliation{%
 \institution{Salesforce Research Asia}
}


\begin{abstract}
Root Cause Analysis (RCA) of any service-disrupting incident is one of the most critical as well as complex tasks in IT processes, especially for cloud industry leaders like Salesforce. Typically RCA investigation leverages data-sources like application error logs or service call traces. However a rich goldmine of root cause information is also hidden in the natural language documentation of the past incidents investigations by domain experts. This is generally termed as Problem Review Board (PRB) Data which constitute a core component of IT Incident Management. However, owing to the raw unstructured nature of PRBs, such root cause knowledge is not directly reusable by manual or automated pipelines for RCA of new incidents. This motivates us to leverage this widely-available data-source to build an Incident Causation Analysis (ICA) engine, using SoTA neural NLP techniques to extract targeted information and construct a structured Causal Knowledge Graph from PRB documents. ICA forms the backbone of a simple-yet-effective Retrieval based RCA for new incidents, through an Information Retrieval system to search and rank past incidents and detect likely root causes from them, given the incident symptom. In this work, we present ICA and the downstream Incident Search and Retrieval based RCA pipeline, built at Salesforce, over 2K documented cloud service incident investigations collected over a few years. We also establish the effectiveness of ICA and the downstream tasks through various quantitative benchmarks, qualitative analysis as well as domain expert's validation and real incident case studies after deployment.
\end{abstract}




\maketitle
\section{Introduction}
\label{sec:intro}
For operating large-scale cloud services, incident (e.g., unplanned interruption, outage or performance degradation) management \citep{10.1145/3368089.3417055} is a critical aspect of IT-Ops, with one of its most complex problems being Incident Root Cause Analysis (RCA) \citep{10.1145/3377813.3381353}. Lately, AIOps (AI for IT Operations) has played a significant role in automating RCA investigation by leveraging data-driven and AI techniques over data-sources like application logs, time series and execution trace data. However in this work, we focus on a rich data-source of past documented incident investigation reports - generally termed as \textbf{P}roblem \textbf{R}eview \textbf{B}oard (PRB) Data \citep{prb,problem_mng} that constitute a core component in all major IT Incident Management pipelines \citep{DBLP:conf/icse/BansalRAMJ20,DBLP:journals/corr/abs-2007-05505,10.1109/DSN.2014.39}.  
\vspace{-0.8em}

\begin{figure}[!bp]
\centering
\vspace{-1em}
\includegraphics[width=0.5\textwidth]{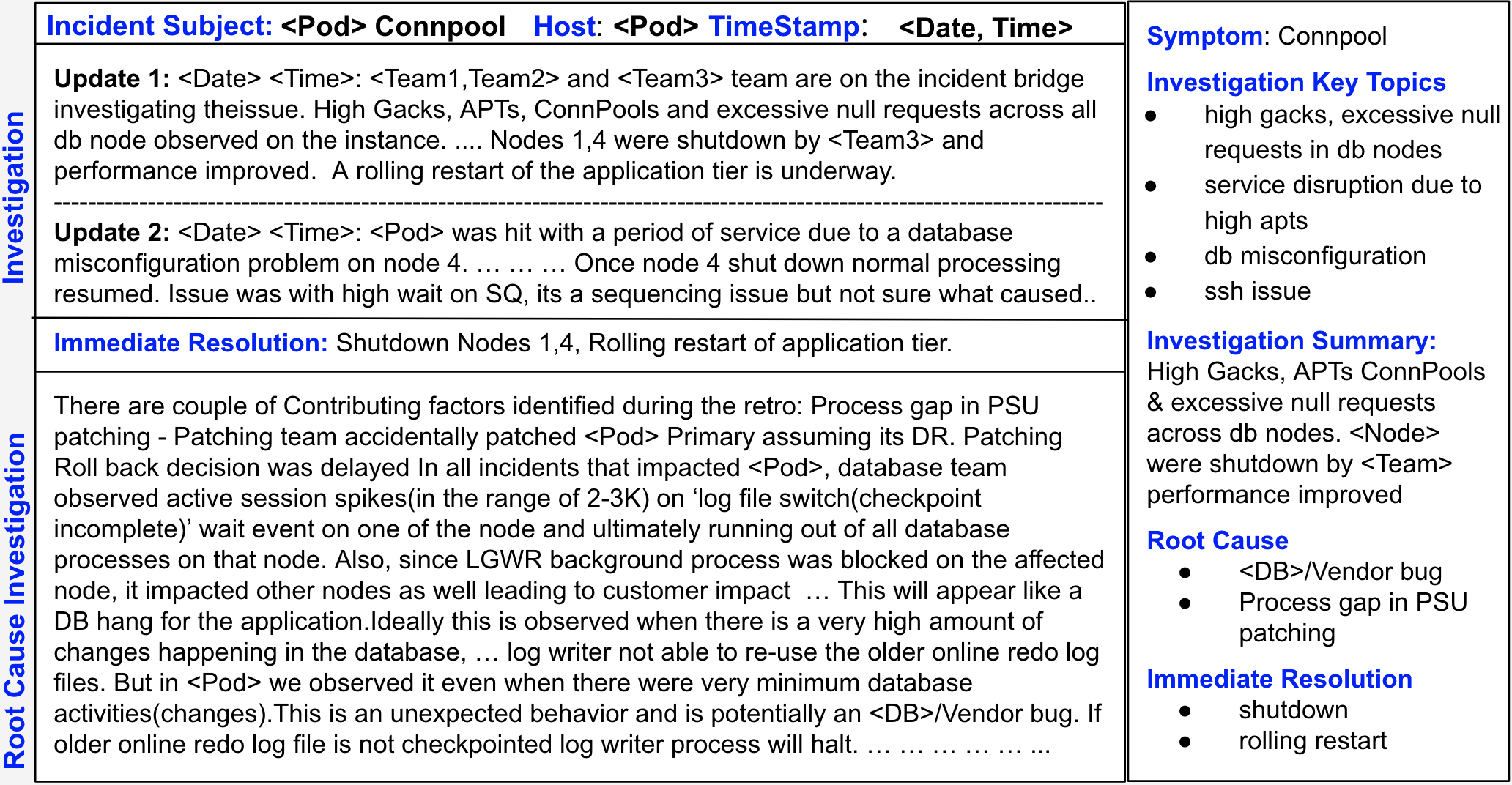}
\caption{(left) Raw PRB document and (right) Structured form obtained through neural information extraction}
\label{fig:struct_know}
\end{figure}
The manual RCA investigation in carried out in 5 steps, with each documented separately in the PRB data:
\textbf{1) Incident Detection} which typically relies on analysis of Key Performance Indices (e.g. Average Page Time of a service), continuously monitored for each host machine.
\textbf{2) Symptom Detection} which detects the primary effect of the service disruption on performance factors like CPU and is treated as the key observable symptom e.g. High App CPU.
\textbf{3) Investigation} which requires multiple iterations of back and forth communication updates between teams of Site Reliability Engineers or SRE inspecting into different aspects to understand the broad nature of the incident and identify the target team who can undertake the RCA investigation. 
\noindent\textbf{4) Immediate Resolution:} Based on the conclusions of the investigation, an action is taken to temporarily mitigate the problem, e.g., an incident due to sudden increase of memory consumption on a node might be mitigated by restarting it, but a deeper investigation needs to be carried out by the respective Service Owner for RCA. 
\noindent\textbf{5) Root Cause Investigation:} Finally the target team for RCA investigates into the true root cause (e.g a bug in the service or a specific process). Each phase of this investigation is documented in a long open-ended form containing evidences from different kinds of systems data (logs, traces etc) or team-communications (email or chat discussions).  This process of documenting incident investigations is indeed a critical component of any major ITOps pipeline, making PRBs a fairly generic data-source largely available in all Incident Management systems. 
Especially with repeated incidents causing similar service disruptions, past incident data has become increasingly valuable. However, in its raw form this rich informative data-source cannot be directly consumed by AI models or domain experts for knowledge reuse, nor are there any existing AIOps pipeline leveraging it.
\begin{figure}[!tp]
\begin{subfigure}{\linewidth}
\centering
\includegraphics[width=\linewidth]{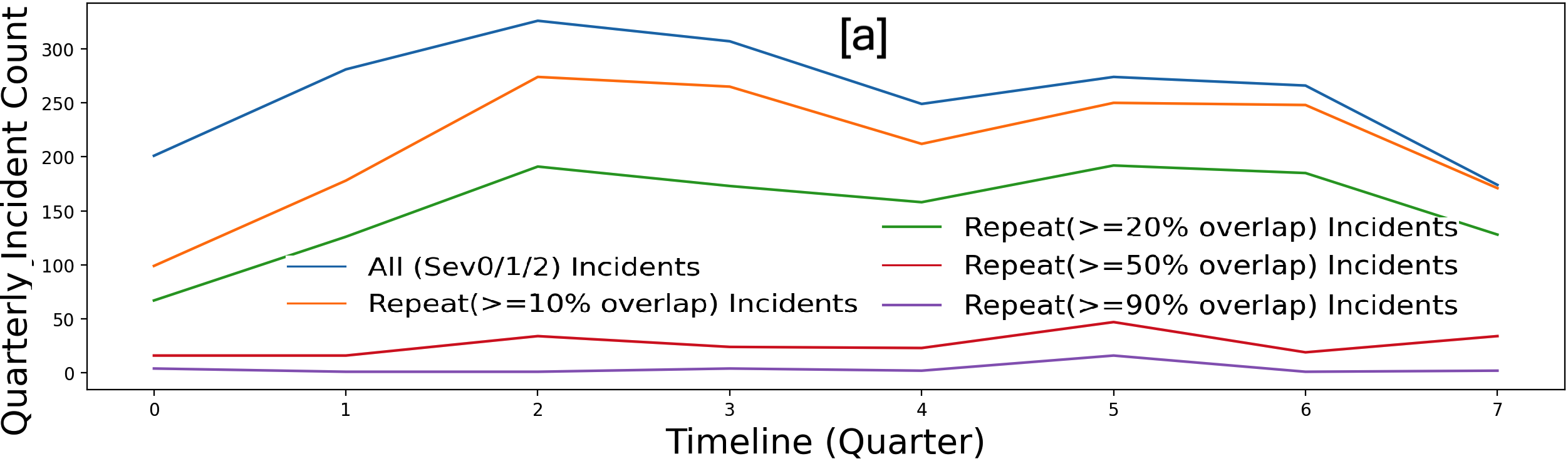}
\end{subfigure}
\\
\begin{subfigure}{0.5\linewidth}
\centering
\includegraphics[width=\linewidth]{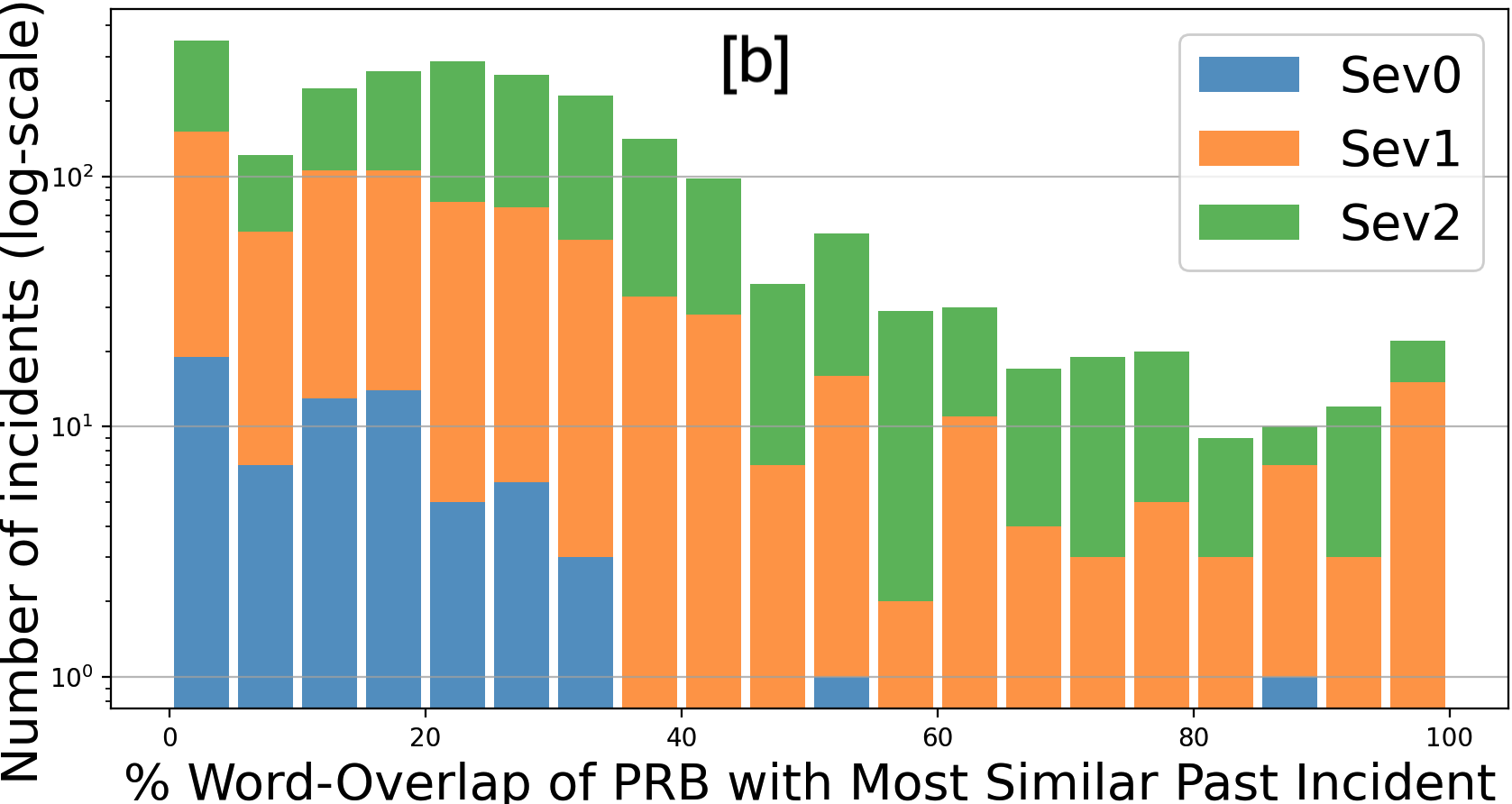}
\end{subfigure}
\begin{subfigure}{0.49\linewidth}
\centering
\includegraphics[width=\linewidth]{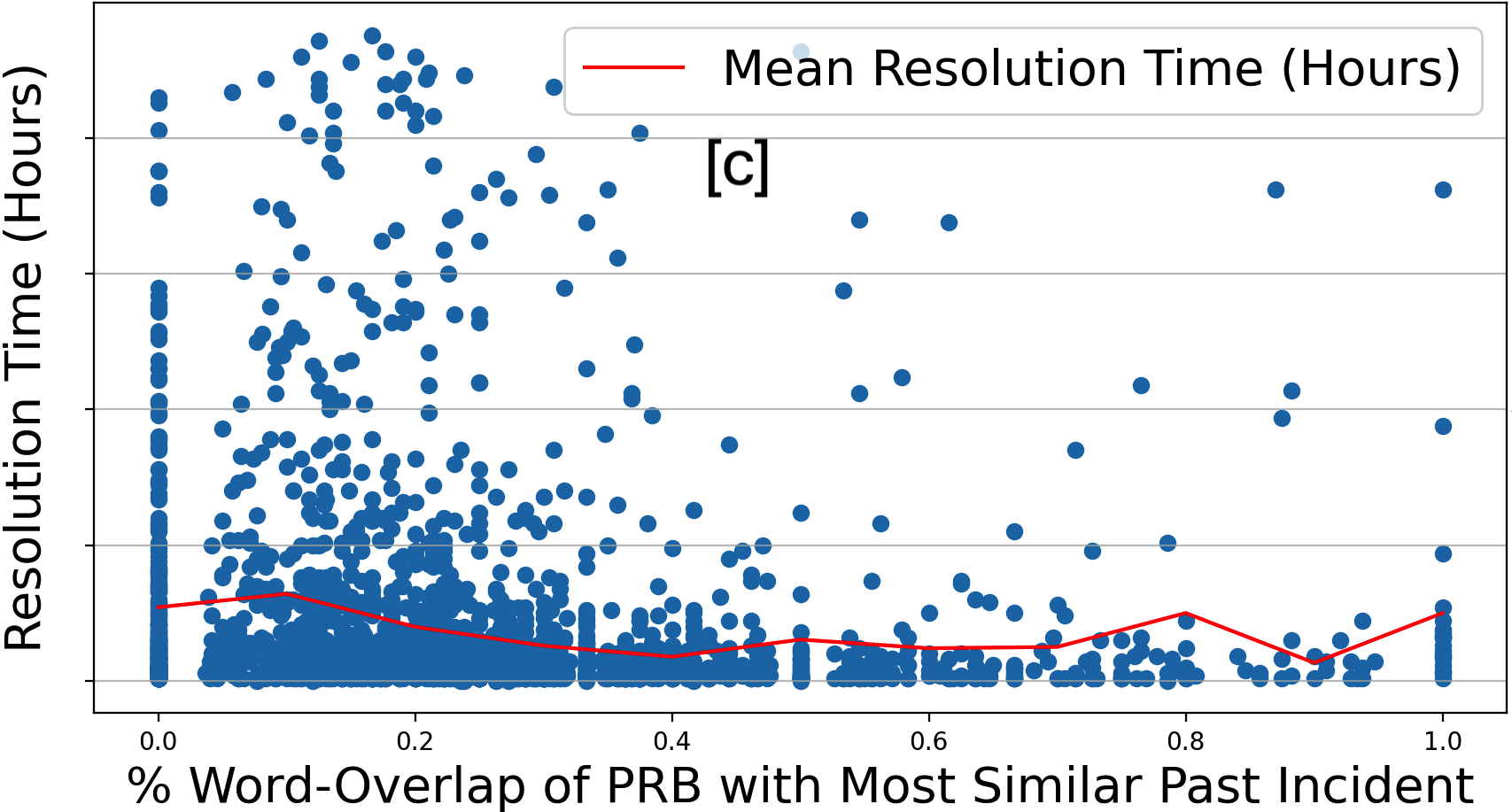}
\end{subfigure}\vspace{-0.5em}
\caption{Plot w.r.t degree of repeatedness of incidents: (a) Timeline of quarterly incident count (b) Distribution of incident severity  (c) Incident Resolution Time}
\label{fig:repeat_incidents}\vspace{-1.5em}
\end{figure}

With this motivation, our main \textbf{Research Question} is \emph{Can AI assist ongoing incident analysis by extracting and mining relevant RCA information from past investigations and reusing this knowledge to quickly resolve repeating incidents?} 

Towards this end, \textbf{Our Primary Contributions} are:
\vspace{-0.3em}
\begin{itemize}[leftmargin=*]
    \item We propose an AI pipeline of Incident Causation Analysis (ICA) over past Incident Investigations (i.e. PRB) data, constituting of 
\begin{itemize}[leftmargin=*]
    \item A targeted Neural NLP and Information Extraction system to extract key RCA information from each unstructured PRB
    \item A specialized Neural Knowledge Mining system to aggregate document-level information over all incidents into a domain-specific Causal Knowledge Graph
\end{itemize}
\vspace{0.3em}
\item Using rich distributional semantics of natural language, ICA enables Incident Search and RCA, given a new incident symptom: 
\begin{itemize}[leftmargin=*]
    \item  A Neural Information Retrieval System to search and rank the most relevant past incidents with similar symptom
    \item A simple yet effective Retrieval based RCA to detect the most likely root causes and resolutions for the given incident symptom based on the past investigations.
\end{itemize}
\end{itemize}
\begin{table}[!tp]
{\footnotesize
    \centering
    \begin{tabular}{|p{7.3cm}|p{.7cm}|}\hline
    \end{tabular}
    \vspace{0.2em}
    \begin{tabular}{|p{6.3cm}|p{1.7cm}|}\hline
        \multicolumn{2}{|c|}{\textbf{Average Stats (mean $\pm$ std) of Raw \& Extracted PRB data fields}}\\ \hline 
        Length (non-stopwords) of PRB Subject Text & 5.2 $\pm$ 2.9\\ \hline 
        Length (non-stopwords) of full PRB Investigation Doc & 309.7 $\pm$ 732.2 \\ \hline
        Length (\ditto) of PRB Resolution Document & 18.3 $\pm$ 24.8 \\ \hline
        Length (\ditto) of PRB RCA Document &  110.1 $\pm$ 159\\ \hline
    \end{tabular}
    \vspace{0.2em}
    \begin{tabular}{|p{6.3cm}|p{1.7cm}|}\hline
        Length (non-stopwords) of extracted Symptom & 4.1 $\pm$ 2.8\\ \hline 
        Length (\ditto) of extracted Investigation Topics & 3.1 $\pm$ 1.0 \\ \hline
        Length  (\ditto)  of extracted Investigation Summary  & 41.6 $\pm$ 31.9\\ \hline 
        Length (\ditto) of extracted Root Cause & 5.9 $\pm$ 4.1\\ \hline 
        Length (\ditto) of extracted Resolutions & 5.0 $\pm$ 3.3\\ \hline 
    \end{tabular}
    \vspace{0.2em}
    \begin{tabular}{|p{2.95cm}|p{1.55cm}|p{1.45cm}|p{1.32cm}|}\hline
        \textbf{Sev Wise Avg Stats} & \textbf{Sev0} & \textbf{Sev1} & \textbf{Sev2} \\ \hline 
        Percentage of Incidents & 3.61\% & 31.23\% & 65.15\%\\ \hline
        Investigation Doc Length & 1479 $\pm$ 2680 & 419 $\pm$ 707 & 193 $\pm$ 300 \\ \hline 
        Resolution Doc Length & 28 $\pm$ 32 & 21 $\pm$ 29 & 17 $\pm$ 21.5 \\ \hline 
        RCA Doc Length & 189 $\pm$ 163 & 152 $\pm$ 150.6 & 86 $\pm$ 157 \\ \hline
    \end{tabular}
    \caption{Stats of our collected dataset of 2000 PRB records}\vspace{-3em}
    \label{tab:dataset_stats}}
\end{table}
Additionally, ICA provides a framework for post-mortem analysis into \emph{all} past incidents and understand the nature of recurring issues. It allows application of formal ML techniques (e.g., causal models or graphical inference) for RCA or, in future, building comprehensive AIOps pipeline for holistic RCA over multimodal multi-source data like logs, execution traces and time series metrics.
By succinctly representing and consuming RCA knowledge of past investigations, ICA provides an ideal framework for novice inexperienced SREs to quickly get onboarded into the RCA process and acquire the knowledge of more experienced SREs. In future, infrastructures like Causal Knowledge Graph, might enable discovering RCA insights even unknown to the experienced SRE. With this objective, in  Sec. \ref{sec:motivation}, we motivate the problem, in Sec. \ref{sec:model} we present the proposed system which we analyze and evaluate in Sec. \ref{sec:eval} and conclude in Sec. \ref{sec:conclusion}.

\section{Problem Setting and Motivation}
\label{sec:motivation}
Our PRB dataset is collected over 2000 investigations of cloud service incidents having Sev0/1/2 severity levels, i.e., respectively, Catastrophic, Critical or High Impact. In this section we motivate the need for mining such a PRB data source and elaborate its challenges.\vspace{-0.5em}
\begin{figure*}[!tp]
\centering
\includegraphics[width=\textwidth]{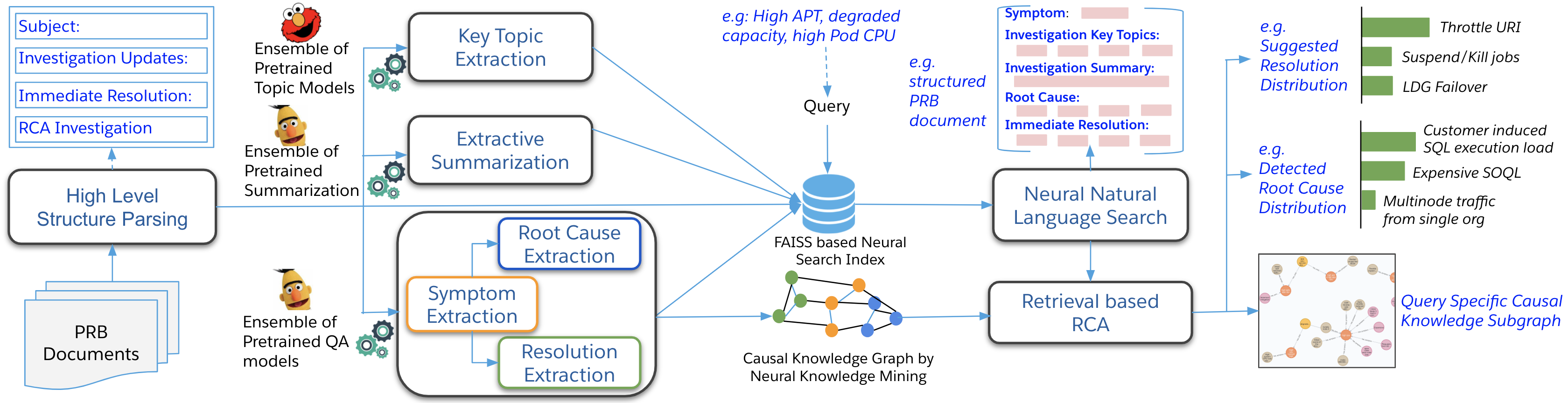}\vspace{-.2em}
\caption{Incident Causation Analysis Pipeline with downstream Incident Search and Retrieval based RCA over PRB Data}
\label{fig:ie_ir_pipeline}

\end{figure*}
\\
\textbf{Motivation-1: Inadequate Prior Work}: Most of RCA literature has focused on mining usable knowledge from Software Logging data and Execution traces, Service Call Graphs and Time Series data. The closest works related to us are \cite{10.1145/3368089.3417054,6982620,DBLP:journals/corr/abs-2007-05505} which mines knowledge from incident descriptions or Troubleshooting Guides or Bug Fixes. However, they focus on more straightforward tasks, involving simpler models or readily available training data. \cite{DBLP:journals/corr/abs-2007-05505} only tags named entities from Incident Descriptions, by leveraging the available structured key-value pair data. \cite{10.1145/3368089.3417054} recommends TroubleShooting guides for incidents with available parallel data to train CNN/LSTM models. \cite{6982620} does bug triaging on structured Bug-Fix documents using non-neural and heuristic methods. Because of their use-cases, they do not require the targeted information extraction and causal mining needed for RCA.
Our other contribution is building a Causal Knowledge Graph from the unstructured incident data. This is a quite different task from the well-studied literature of causal graph construction \cite{10.1007/978-3-030-03596-9_1,DBLP:journals/corr/abs-2007-05505,BRANDON2020110432} from structured call traces/graphs and system dependency graphs. \vspace{0.5em}
\\
\noindent\textbf{Motivation-2: PRB as Oracle Root Cause Knowledge Source:}
Traditional RCA pipeline heavily rely on data-sources like application or error logs \citep{10.1145/3338906.3338961,10.1145/3038912.3052649,10.5555/2228298.2228334,10.1145/3133956.3134015,5609556} from service health monitoring tools or execution traces \citep{10.1007/978-3-030-03596-9_1,8411065,2021arXiv210301782L,8711092} and time series of KPI metrics \citep{app10062166,7563819,DBLP:journals/corr/abs-2007-05505,7820614}. However discovering any root-cause related signal in these data-sources can be a complex time-consuming task. On the other hand, the past incident investigations documented by domain experts are a rich goldmine of \emph{Oracle} Root Cause Information, containing many explicit informative linguistic cues connecting the incident symptom to the expert-recommended root cause and resolution. While in its raw form, these long unstructured documentations are not apt for knowledge reusal, SoTA NLP techniques can process them into a structured form amenable to AI models. \vspace{0.5em}
\\
\noindent\textbf{Motivation-3: Handling Repeated Incidents:}
A repeated incident is one with similar symptom, root cause and resolution as any past incident. We quantitatively define extent of repeatedness as the maximum obtainable Word-Overlap of the concatenation of these three fields, when compared with all \emph{past} incidents. Fig. \ref{fig:repeat_incidents}(a) plots the timeline of quarterly count of all and various degrees of repeated incidents, showing that the latter consistently persists throughout the period. In Fig. \ref{fig:repeat_incidents}(b), we observe that the distribution of incident severity (Sev1/Sev2) is quite similar across repeated and non-repeated incidents, indicating that repeated incidents need as much attention as the non-repeated ones. And Fig. \ref{fig:repeat_incidents}(c) shows that the distribution of incident resolution time (and its mean) is quite similar across repeated and non-repeated incidents. This is largely due to the lack of a framework of knowledge-reusal from past investigations. Especially with increasing number of high-stake recurring incidents, AI pipelines have become essential to extract and represent the RCA knowledge embedded in PRBs.\vspace{0.5em}
\\
\noindent\textbf{Challenge-1: Long Unstructured Documentation of PRB}
We need to handle the long, unstructured content in PRBs that are often rife with irrelevant distractive information. Table \ref{tab:dataset_stats} presents the statistics of the original PRB documents as well as the extracted information. For instance, it shows that more severe incidents  (Sev0$>$1$>$2) typically result in longer documentation.\vspace{0.5em}
\\
\noindent\textbf{Challenge-2: Unsupervised Setting} We need to tackle this complex domain-specific problem in an unsupervised setting, using generic pretrained or unsupervised NLP models, with minimal assumption on the nature of PRB documents. \vspace{0.5em}
\\
\noindent\textbf{Can PRBs be organically generated in a structured form?} RCA investigations typically require multiple iterations of cross-team communications. This decentralized documentation in an agile troubleshooting and triaging framework results in PRB fields being direct excerpts from email-exchanges or chat-conversations, making it impossible to generate PRBs in a structured form. 
\\

\vspace{-1em}
\section{Our Approach}
\label{sec:model}
The raw PRB documents (from Fig. \ref{fig:struct_know}) recording incident investigations have a generic structure comprising of: 
i) Incident Subject, i.e., a title capturing the primary symptom of the incident, ii) Incident TimeStamp and Host Details, iii) Sequence of updates of the Investigation, iv) Immediate Resolution and v) Root Cause Investigation. With this form of PRB Data, Fig \ref{fig:ie_ir_pipeline} shows our proposed Incident Causation Analysis (ICA) pipeline (which we elaborate below) whose main components are: \\i) Offline Neural Information Extraction from unstructured data \\ii) Neural Knowledge Mining to construct Causal Knowledge Graph \\iii) Neural Search to find past investigations relevant to any query\\iv) Retrieval based RCA using Neural Search \& Causal Knowledge Graph to predict root causes and resolutions for a query symptom.
\vspace{-0.2em}
\subsection{Neural Information Extraction}
\label{subsec:ie}
Here we extract targeted information from the unstructured PRBs.\vspace{0.3em}
\\
\noindent\textbf{Symptom Extraction:} This involves a rule-based module to extract the generic symptom indicating the incident (e.g., connpool) from the PRB Subject, by removing specific Host Machine details. At inference time, for a new incident, the symptom is observed from the KPI metrics and provided as input to the ICA pipeline. \vspace{0.5em}
\\
\noindent\textbf{Key Topic Extraction:} For this we employ an ensemble of  various unsupervised Topic models \citep{boudin:2016:COLINGDEMO} to extract short topical phrases most representative of the document. This includes Graphical Topic Models like TextRank, SingleRank, TopicRank, TopicalPageRank, PositionRank, MultipartiteRank, Feature based model YAKE, Neural embedding based SIFRank which uses pretrained ELMO. The extracted phrases along with their probability scores are simply aggregated into a distribution by the ensembling technique.\vspace{0.5em}
\\
\noindent\textbf{Root Cause \& Resolution Extraction:}
To extract targeted spans containing crisp root causes and resolution information from the PRB document, we use pretrained Transformers that have been fine-tuned on Machine Reading Comprehension task that answers questions based on a given passage. Specifically we apply an ensemble of BERT models (DistilBERT, BERT-base and BERT-large)\citep{devlin-etal-2019-bert}, RoBERTa \citep{liu2019roberta} and SpanBERT \citep{joshi-etal-2020-spanbert}, each fine-tuned on the standard open-domain extractive-QA dataset SQuAD. For extracting root causes, we create a hand-crafted query-template, e.g., ``\emph{What was the root cause of SYMPTOM?} or \emph{What caused the incident?}", where \emph{SYMPTOM} refers to the extracted incident symptom. Similarly for resolution extraction we use queries like ``\emph{What was done to remedy the SYMPTOM?} or \emph{What action resolved the incident?}". Empirically, we observe that these pretrained QA models perform quite well on this challenging problem, without any supervision of our target AIOps domain. These models extract short spans with a probability score, which are aggregated by an ensemble after lexical de-duplication resulting in arbitrarily many target spans.\vspace{0.5em}
\\
\noindent\textbf{Extractive Summarization:} In order to extract the most informative sentences from the PRB investigation as a summary, we use a RoBERTa \citep{liu2019roberta} based extractive summarization model finetuned on a standard benchmark summarization dataset, CNN-DailyMail.

\subsection{Neural Knowledge Mining}
\label{subsec:km}
Our Neural Knowledge Mining module aggregates the document-level information extracted in form of symptoms, root causes and resolutions and builds a global Causal Knowledge Graph (CKG). This infrastructure not only enables visualization of the causal structure underlying the incidents, it also plays a key role in predicting the likely root cause and resolution given a symptom. We store the CKG in the popular GraphDB framework, Neo4j (\url{https://neo4j.com}) which allows interactive visualization and navigation and efficient querying over the graph. Below are the steps to construct the graph:\vspace{0.5em}

\noindent\textbf{PRB Document Level Graph:} For each document, nodes are constructed out of the extracted symptom, root-cause and resolutions and edges added individually between symptom, root-cause and symptom-resolution pairs. Each node is represented as a vector obtained as average of the GloVe \citep{pennington-etal-2014-glove} based token embeddings of the node description, weighted by normalized term-frequency.\vspace{0.5em} 

\begin{figure}[!tp]
\centering
\includegraphics[width=\linewidth]{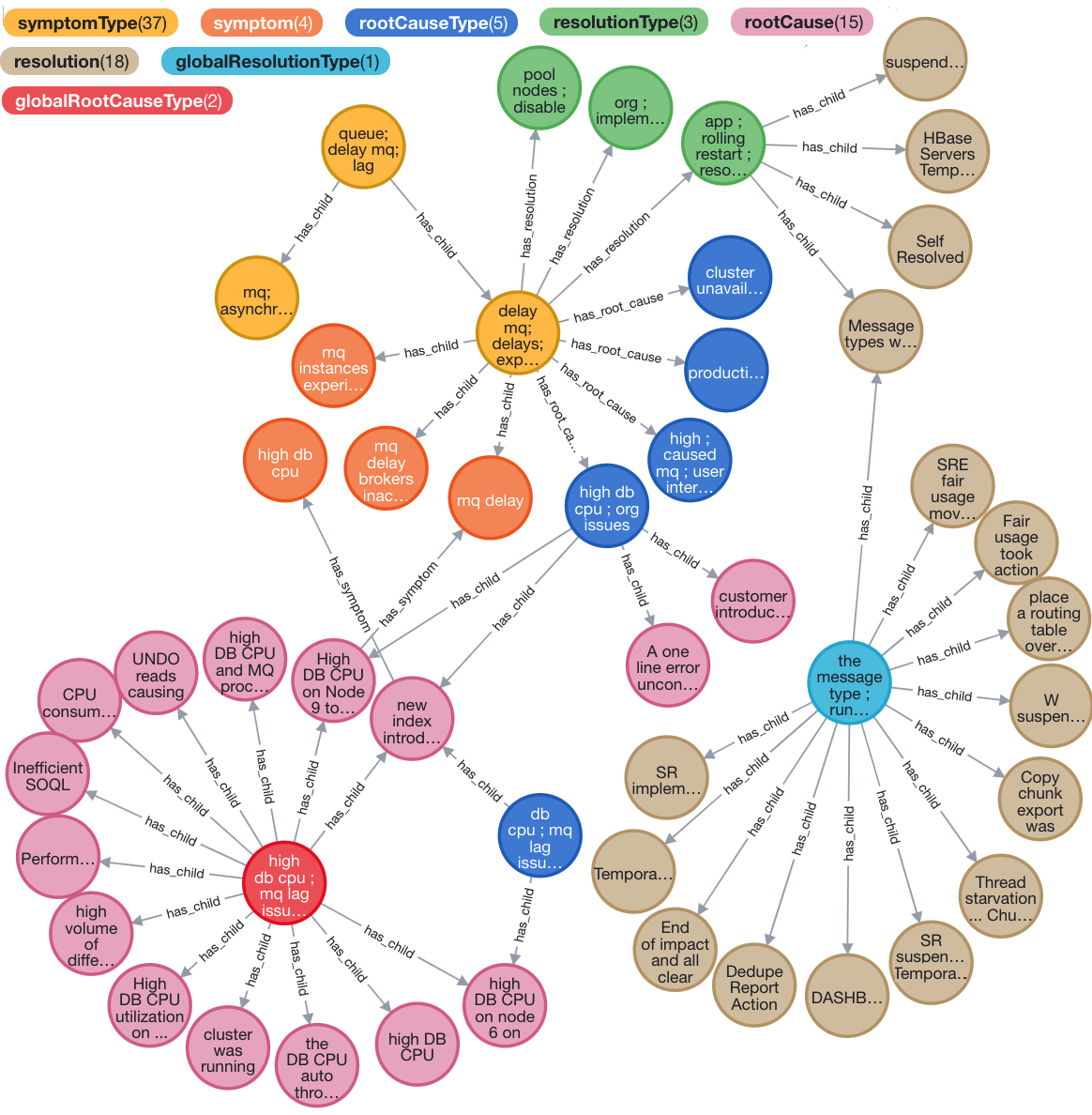}
  \caption{Snippet of constructed Causal Knowledge Graph}
\label{fig:causal_graph}\vspace{-1em}
\end{figure}

\noindent\textbf{Aggregating Document Level Graphs:} We apply simple clustering strategies over the dense representation of the graph nodes to aggregate the incident level information into a compact Causal Knowledge Graph. Fig. \ref{fig:causal_graph} shows a randomly selected subgraph, with the different node-types, each with label as the corresponding textual description extracted from the PRB document or the generated cluster label. The clustering involves the below steps:\vspace{0.5em}

\noindent\textbf{Step 1: Clustering Symptoms:} We apply Hierarchical Agglomerative Clustering techniques \citep{scikit-learn} to hierarchically cluster the symptoms extracted from all the past PRB records, bottom to top, into a \emph{symptomType} node, successively merging them together by minimising the sum of squared distances within all clusters. Number of clusters are estimated using ELBOW and Silhouette methods.\vspace{0.5em}

\noindent\textbf{Step 2: Clustering Root Causes and Resolutions:} Next, we individually apply \emph{global} and \emph{local} Affinity Propagation \citep{scikit-learn} based clustering on the root-causes and resolutions through iterative message passing between nodes as functions of their similarity score. Here \emph{global} refers to clustering over all root-causes (and resolutions) collected from all PRB records, leading to \emph{globalRootCauseType} and \emph{globalResolutionType} nodes as cluster-heads. And, \emph{local} clustering is only restricted to the root-causes (and resolutions) associated to symptoms within each cluster of \emph{symptomType}, creating cluster-heads \emph{rootCauseType} and  \emph{resolutionType} for that symptom cluster.\vspace{0.5em}
\\
\noindent\textbf{Step 3: Cluster Labeling:} For each cluster, we construct a document by concatenating all the node labels and use a popular collocation extraction technique \citep{1708523} on it. It greedily selects the longest n($\leq$3)-grams with highest Pointwise-Mutual Information from the document. The selected n-grams are further reranked using average normalized term-frequency weights of the non-stopwords.
\subsection{Using Causal Knowledge Graph for RCA}
\label{subsec:ckg}
The constructed graph has many uses in RCA: \\
\textbf{Global Incident Analysis:} It provides interactive visualization and intuitive navigation over the underlying structure of symptoms, root causes and resolutions connecting different incidents. It also serves as post-mortem analysis tool, to understand the nature of symptoms and root causes of the most recurring past incidents. \vspace{0.5em}
\\
\textbf{Incident-specific Analysis:} Starting with an incident symptom or root-cause, a user can extract the relevant subgraph of the Causal Knowledge Graph or navigate the local cluster of related root-causes or similar symptoms. \vspace{0.5em}

\noindent\textbf{Root Cause or Resolution Prediction from Symptom:} Our primary use-case of the CKG is to leverage the graphical structure to collectively infer the root cause or resolution from \emph{all} relevant past investigations given an incident symptom. We frame them as two independent tasks: i) Root Cause ii) Resolution Prediction, given the Symptom, for which we propose a simple yet effective graphical neural model. For the first task we consider the CKG over global clustering of the symptom and root-cause nodes and ignore the resolution nodes and vice-versa for the second task. We describe the model for the first task which similarly follows for the second one. For simplicity, we ignore the hierarchical clustering over symptoms and only consider the final level of symptom clustering, i.e. \emph{symptomType} here refers to the leaf level cluster heads. \vspace{-1.2em}
\begin{itemize}[leftmargin=*]
    \item $V_S, V_R$ = Set of \emph{symptom} and \emph{root-cause} nodes respectively
    \item $E_S = \{(v_i,v_j) | v_{i},v_{j} \in V_S, cluster(v_i)=cluster(v_j)\}$
    \item $E_R = \{(v_i,v_j) | v_{i},v_{j} \in V_R, cluster(v_i)=cluster(v_j)\}$
    \item $E_{SR} = \{(v_s,v_r) | v_s\in{V_S}, v_r \in V_R, incident(v_s)\cap incident(v_r)\neq \{\}\}$ i.e. they both belong to the same incident
\end{itemize}
\vspace{-0.2em}
This results in a bipartite graph, between the cluster graph formed over the symptom nodes and a similar one formed over the root-cause nodes. During training and inference we assume that the cluster graph over the symptoms and root causes (i.e. $V_S \cup V_R, E_S, E_R$) is completely known. However, the ``symptom root-cause edges'', $E_{SR}$, which are observations from past incidents are unknown for some or all incidents. Our practical motivation is that while organizing known symptoms and root-causes into clusters can be done apriori, there needs to be ``collective learning'' over past incidents to predict the root-cause given a new symptom. Accordingly we frame the problem of predicting Root Cause for a given Symptom, as a Link Prediction task between the symptom and root-cause nodes, given the cluster graphs over symptoms and root-causes. Empirically we show that by training with the knowledge of the symptom root-cause connection from only 1\% of incidents, our model achieves an appreciable link prediction performance on the test incidents.\vspace{0.5em}  
\\
\textbf{Data Augmentation:} During training we further add noisy edges $E^{n}_{SR}$ between symptom and root-cause nodes, by connecting their one-hop neighbours, obtained from the respective cluster graphs. For both tasks, our number of true edges $|E_{SR}|$ constitutes of only 5\% of all (i.e. true and noisy i.e. $|E_{SR}|+|E^{n}_{SR}|$) edges. \\
$E^{n}_{SR} = {(v_i,v_j) | \exists (v_i,v_k) \in E_S, (v_l,v_j) \in E_R, (v_k,v_l) \in E_{SR}}$. \vspace{0.5em}
\\
\textbf{Training Data:} We train different models by varying how the training and validation data is selected. We follow two setups i) All Edge Sampling ii) Noisy Edge Sampling. Respectively for them, we sample $x$\% of edges from All edges i.e. $(E_{SR} \cup E^{n}_{SR})$ or Noisy edges i.e. $E^{n}_{SR}$ for the Train+Validation data split. In both cases, we build multiple models by varying $x$ $\in$ \{1\%,2\%,5\%,10\%,20\%\}.\vspace{0.5em}
\\
\textbf{Model:} We use a Graph Convolutional Network (GCN) \cite{DBLP:conf/iclr/KipfW17} based neural model over the Symptom Root-Cause bipartite graph. GCN has been popularly used for learning representations over graphs, by iteratively updating node representation with their neighbours. First, we compute symptom and root cause node representations as average of GloVe based token embeddings (with dimension $d$). $\tilde{X}_S \in R^{|V_S|\times d}$ and $\tilde{X}_R \in R^{|V_R|\times d}$ represent the node embedding matrices of Symptom and Root Cause nodes. Then we add separate GCN layers on the cluster graphs over symptoms and root causes. This results in updated representations of symptom and root-cause nodes by message passing over their immediate neighbours.\\ 
$X_S = Linear(ReLU(GCN(\tilde{X}_S, E_S)))$\\
$X_R = Linear(ReLU(GCN(\tilde{X}_R, E_R)))$\\
For each training instance, we sample an edge $(v_s, v_r)$ from the training data and randomly sample $n$ negative root causes $\{v^{-}_{r,j} | j=1,\cdots,n, (v_s,v^{-}_{r,j}) \not\in All Edges\}$ and compute their GCN transformed representation i.e. $\{x_s, x_r, \{\cdots x^{-}_{r,j}\cdots\}\}$. Then for the positive pair and each of the negative pairs of symptom-rootcauses, we compute a dotproduct similarity between the symptom and the respective root cause representation.  We then train the neural network with a negative log likelihood of the positive pair of symptom-rootcause. 
\vspace{-0.5em}
\begin{equation*}
    L(v_s, v_r, \{\cdots v^-_{r_j} \cdots\}) = - log (\frac{e^{sim}(x_s,x_r)}{e^{sim}(x_s, x_r) + \displaystyle\sum_{j=1}^n e^{sim}(x_s, x^{-}_{r,j})})\vspace{-0.3em}
\end{equation*}
During inference, given a symptom, we compute the dotproduct of the GCN transformed representation with that of each rootcause node in the constructed CKG and output the one which yields the highest similarity score as the predicted root cause.
\subsection{Incident Search and Retrieval based RCA} 
\label{subsec:search}
We now focus on ICA modules catering to our RCA goal. After a new incident occurs and its key symptoms have been identified, a core Incident Management task is to efficiently search over past similar incident investigations and promptly detect the likely root causes and remediations based on them. For this we build a specialized Incident Search and Retrieval based RCA which takes any open-ended natural language query and 
i) retrieves the most similar past incidents and presents the key RCA information in a structured form 
ii) predicts the most likely root causes and resolutions for the queried symptom based on the past investigations.\vspace{0.5em}
\\
\textbf{Incident Search:} Our Incident Search engine is based on neural search which represents and indexes documents as dense high-dimensional vectors and retrieves them using efficient vector-space search. Neural search has several advantages over traditional symbolic search, like seamlessly handling typological errors, domain specific abbreviations or agglutinations and non-canonicalized wordings abundant in manual documentations (e.g. ``connection pool'' as ``conpool'' or ``conn pool'')
\vspace{-0.2em}
\noindent\begin{itemize}[leftmargin=*]
\item \textbf{Data Representation:} The search index is built over the data from the raw PRB Subject and Investigation document as well as all the targeted information extracted from them. Each topic or sentence from the raw documents or extracted summaries is represented as a vector, obtained as the average of the contextualized token embeddings from the pretrained RoBERTa \citep{liu2019roberta}.\vspace{0.3em} 
\item \textbf{Indexing:} Each sentence or phrase is separately indexed as vector by FAISS \citep{JDH17} which allows fast approximate nearest-neighbor retrieval over high dimensional vector space. \vspace{0.3em} 
\item \textbf{Query Representation:} Queries being typically short, are also represented as average of RoBERTa based token embeddings.\vspace{0.3em} 
\item \textbf{Retrieval \& Ranking:} FAISS search retrieves the most relevant sentences over all PRB Documents, scoring them w.r.t query based on standard vector similarity metrics (e.g., dot-product). These sentence-level scores are max-pooled to get an overall document-level score for each of the top-K retrieved results.\vspace{0.3em}
\item \textbf{Visualizing:} The retrieved PRB documents are then shown in an easily consumable structured form of the extracted information:  Investigation Subject, Topics, Summaries, Root Cause and Resolution. Apart from this, we also show a query specific subgraph of the CKG associated with the top-k retrieved results. This allows users to interactively navigate over the related incidents and explore their underlying root cause and resolution clusters.\vspace{0.3em}
\end{itemize}
\textbf{Retrieval based RCA:} We apply the following two techniques for obtaining a distribution of most likely root causes and resolutions (which follows similarly) from the top-K retrieved incidents.
\noindent\begin{itemize}[leftmargin=*]
\vspace{-0.2em}
\item \textbf{Using Incident Search:} For each search result, the multiple extracted root cause spans are merged into a single sentence-form with the span scores max-pooled and multiplicatively combined with the relevance score of the search result. After de-duplication of the root-causes and score-normalization we obtain the distribution over the top-k likely root-causes.\vspace{0.3em}
\item \textbf{Using Incident Search and Causal Knowledge Graph:} We take the symptom extracted from the top-k retrieved search results and for each symptom we predict the top-k likely root-causes using the Causal Knowledge Graph as described in \ref{subsec:ckg}. The root cause prediction score is multiplicatively combined with the search result relevance score, and aggregated over repeated occurrences, if any. The resulting top-k root causes form the predicted distribution with their final scores L1-normalized.\vspace{0.3em}
\end{itemize}

\vspace{-1em}
\begin{table}[!htbp]
{\footnotesize
    \centering
    \begin{tabular}{|p{0.5cm}|p{0.4cm}|p{1.3cm}|p{0.9cm}|p{1cm}|p{0.8cm}|p{1cm}|} \hline
    \textbf{Info} & \textbf{Total}  & \multicolumn{5}{|c|}{\textbf{Annotation Labels \& Overall Metrics}} \\ \hline 
    \textbf{Topics} & 1320 & WellFormed: 1276  & Informat-ive:1009 & Uninteres-ting:248 & Unclear: 60  & Extra Words: 15 \\ \hline 
    \textbf{Summ-ary} & 525 & Informative: 435 &  \multicolumn{2}{|p{1.8cm}|}{Too Specific: 46} & \multicolumn{2}{|p{1.9cm}|}{Too Generic: 43} \\ \hline 
    \textbf{Root Cause} & 320 & ExactMatch: 79.07\% & \multicolumn{2}{|p{2cm}|}{F1 (Bag of NonStop Words): 87.97\%} & \multicolumn{2}{|p{1.7cm}|}{F1 (Bag of All Words): 88.44\%} \\ \hline
    \textbf{Resol-ution} & 175 & ExactMatch: 70.34\% & \multicolumn{2}{|p{2cm}|}{F1 (Bag of NonStop Words): 81.57\%} & \multicolumn{2}{|p{1.7cm}|}{F1 (Bag of All Words): 81.69\%} \\ \hline
    \end{tabular}
    \caption{Human validated results of Information Extraction}
    \label{tab:survey}}
    \vspace{-.5em}
\end{table}
\vspace{-1em}
\section{Evaluation and Analysis}
\label{sec:eval}
In this section we evaluate the ICA modules and  downstream Incident Search and RCA tasks over our in-house collected PRB dataset of 2000 incidents. As the first work to leverage PRB data for AIOps and RCA, our objective is to validate the feasibility of using generic NLP models in mining PRB data for building a specialized tool like ICA. We present empirical results through systematic quantitative evaluations, qualitative analysis and surveys,  expert-annotated validation of the model predictions, and real incident case studies after deployment in which we seek the SRE’s judgement on the effectiveness of the proposed model’s predictions in our approach.
\subsection{Neural Information Extraction}
\label{subsec:eval_ie}
\textbf{Human Validation of Extracted Information:} In Table \ref{tab:survey} we provide the quantitative results of a survey conducted over domain experts and target end-users. Its aim is to validate and annotate the quality of topics, summaries, root cause and resolutions that are extracted from PRB documents (Sec. \ref{subsec:ie}).\vspace{0.5em}
\begin{figure*}[!htbp]
\centering 
\includegraphics[width=\textwidth]{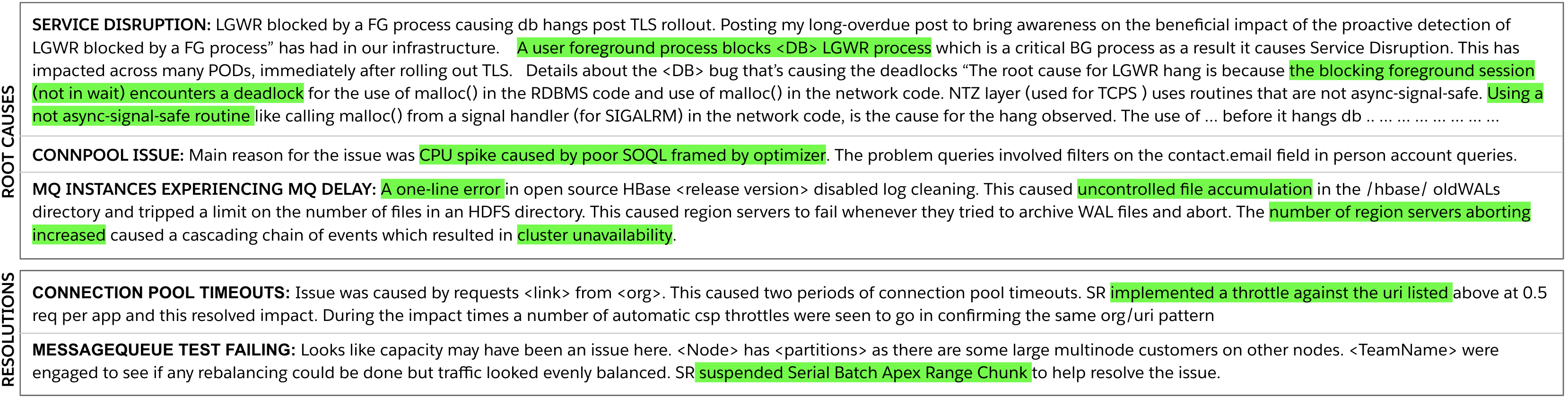}\vspace{-1em}
\caption{Examples highlighting the (top) Root Cause and (bottom) Resolution spans respectively extracted from the Root Cause and Immediate Resolution data fields in PRB documentation (with corresponding PRB Subject in bold)}
\label{fig:rc_example}
\end{figure*}
\\
\noindent\textbf{Topics:} We selected 1320 topics over all PRB documents, sampling uniformly from the topic score distribution. On this, annotators were asked to provide following (binary) labels: i) Grammatically Well-formed (may not be informative) ii) Sufficiently Informative iii) Clarity in Meaning iv) Too Generic or Uninteresting iv) Has extra irrelevant words. As the results show, most of the topics are well-formed and around 76\% are found to be informative and useful.\vspace{0.5em}
\\
\noindent\textbf{Summary:} We similarly take the generated summary of 525 PRB documents and ask the annotators to provide the following binary labels i) Satisfactorily Informative ii) Too Specific (i.e., has additional irrelevant sentences) iii) Too Generic (e.g., does not have any information about the outcome of the investigation). However, sometimes, the summary is too generic due to the original PRB document being incomplete. Despite that, around 83\% of summaries are found to be informative with appropriate level of detailing. \vspace{0.5em}
\\
\noindent\textbf{Root Cause and Resolutions:} We provide the annotators with randomly sampled 320 Root Cause documents and 175 PRB Resolution documents, respectively with their extracted root cause and resolution spans highlighted in them. The annotator is asked to freely modify or delete any span that is found to be not grammatically well-formed or incorrect as the root cause or resolution. The annotator can also independently add other spans deemed as correct. The overall results show that the unsupervised models indeed perform remarkably well. 79\% of the predicted root-cause and 70\% of resolution spans are found to be \emph{exactly} correct and the (micro) average F1-Score of the predicted and annotated spans (in terms of Bag of All Words or Bag of Non-StopWords) is around 88\% and 81\% respectively for root causes and resolutions. \vspace{0.5em}

\noindent\textbf{Qualitative Analysis of Root Cause \& Resolution Extraction}: In Fig. \ref{fig:rc_example}, we illustrate a few examples of the root cause and remedial actions respectively extracted from the raw Root Cause and Immediate Resolution data fields of past incidents. We observe that i) our models are able to extract relevant root cause or resolution spans, despite the long unstructured nature of the document having abundant unseen technical jargon ii) the model does not show any undesirable bias towards passage location in span selection iii) the selected spans are well-formed short crisp self-explanatory topics.
\vspace{-0.1em}

\subsection{Neural Knowledge Mining}
\label{subsec:eval_km}
Next we present some qualitative analysis and illustrative examples of the Neural Knowledge Mining component.

\begin{figure*}
 \begin{minipage}{0.72\textwidth}
 \centering
{\scriptsize
    \begin{tabular}{|p{3.2cm}|p{0.35cm}||p{3.1cm}|p{0.35cm}||p{3.1cm}|p{0.35cm}|}\hline 
    \textbf{Symptom Clusters} & \textbf{\%} &  \textbf{Root-Cause Clusters} & \textbf{\%} & \textbf{Resolution Clusters} & \textbf{\%} \\ \hline
connection pool, high, active requests & 20.65 & the db cpu, materialized view logs, hangs post tls & 4.7 & app tier, ops sfdc net, appeals restore access & 7.41 \\ \hline
db nodes, ops sfdc net mq, high cpu lag & 6.56 & a bug in, packet loss latency, db psu & 3.31 & issue self resolved, auto throttle, rebalanced balancing & 7.08 \\ \hline
sandbox service disruption, service failure, edge services & 5.89 & blocks db lgwr, wait encounters deadlock, async signal safe & 3.18 & self resolved, high db cpu, active session & 5.09 \\ \hline
org migration, intermittent conn pools, service shard, mq sfdc & 5.5 & high memory, redo generation, concurrency issue & 3.05 & app tier, rolling restart, restarted apps & 4.89 \\ \hline
service performance degradation, message queue & 5.16 & on the app, conn pool errors, custom lwc component & 3.05 & conn pool, disabled node, was restarted & 4.7 \\ \hline
message queue processing, refocus test, thread starvation & 4.15 & logfile switch, writer process waiting, checkpoint incomplete & 2.73 & requests sec, bounced broker, restart of & 4.1 \\ \hline
    \end{tabular}
}
\end{minipage}
\begin{minipage}{0.27\textwidth}
\begin{tabular}{@{}c@{}}
    \includegraphics[width=\linewidth]{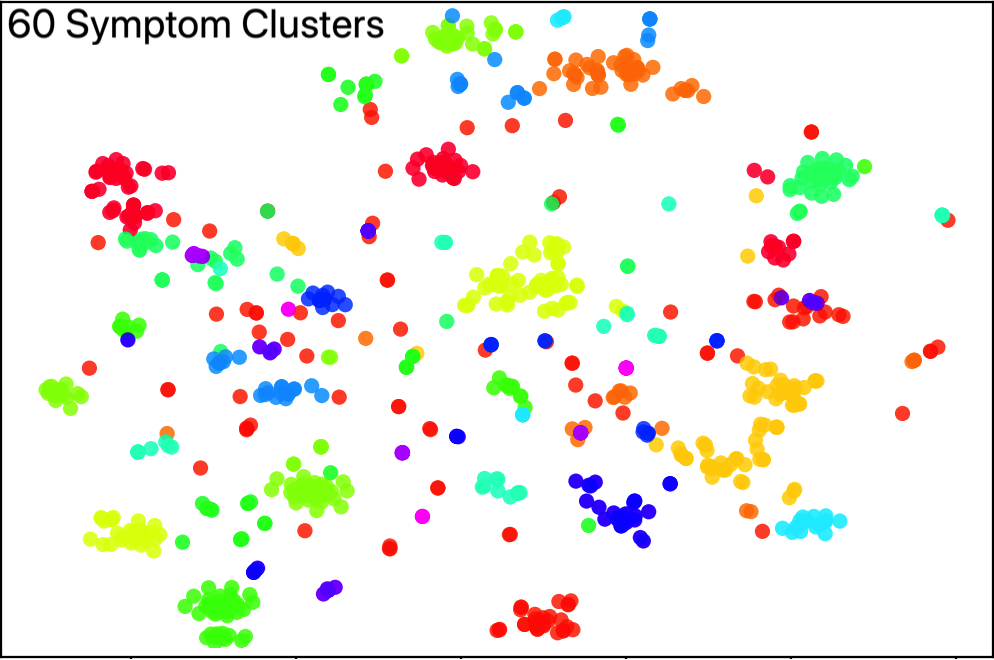}
  \end{tabular}
  
\end{minipage}
\vspace{-0.1in}
\caption{(Left) Distribution of top clusters of Symptoms, Root Causes, Resolutions in terms of \% Coverage of Incidents (Right) t-SNE visualization of the 60 Symptom clusters}
\label{fig:clustering}
\end{figure*}

\noindent\textbf{Symptom, Root-Cause and Resolution Clustering:}
For clustering, we collect unique descriptions of 1072 symptoms, 1915 root causes and 2250 resolutions extracted over all PRB documents. Fig. \ref{fig:clustering} (right) presents the t-SNE visualization (using the node embedding vector) of the resulting 60 Symptom clusters, showing that the clustering is indeed quite well-separated and distinctive. \vspace{0.1em}
\vspace{0.2em}
\noindent\textbf{Post-Incident Analysis:} The Causal Knowledge Graph built over the PRB repository can be instrumental in an organization-wide post-mortem analysis of all past incidents. For e.g. it can analyze the categories of most commonly occurring symptoms or root-causes and resolutions in the last month. In Fig. \ref{fig:clustering} (left) we present the generated cluster labels of the most frequently occurring clusters of symptoms, root causes and resolutions, along with the \% coverage of incidents. This also illustrates the quality of the \textbf{semantic labeling of clusters}. Even with simple greedy n-gram selection techniques (Sec. \ref{subsec:km}), the generated cluster labels still constitute of reasonably coherent well-formed and self-explanatory topics. 
\vspace{-0.5em}
\subsection{Incident Search and Retrieval based RCA}
\label{subsec:eval_search}
In this section we evaluate the performance of the proposed ICA pipeline in retrieving past incidents with similar symptoms and predicting the likely root causes and resolutions from those investigations. Since the PRB data itself is the only Oracle data available to us, we showcase this based on the below quantitative benchmarking technique. In our experimental setup, say, an incident from a given PRB document is \emph{the} ongoing target incident, with its only known information being the incident symptom. Then assuming \emph{all} the remaining PRB documents are available to search and analyze over, our benchmarking is targeted at two questions\vspace{0.1em}\\
\textbf{Q1:} Is Incident Search able to retrieve PRB documents having a high overlap with the target PRB document, not only in terms of the original investigation, but also the targeted extracted information?\vspace{0.2em} \\
\textbf{Q2:} Does the root cause/resolution predicted by ICA match the true root cause/resolution extracted from the target PRB document?\vspace{0.3em}

\noindent For each PRB instance, we take the incident symptom as query and search over the repository of remaining PRB records and compare the similarity of the retrieved top-k results with the target PRB. There being no gold standard of search result set, our metrics are precision-based. We decompose the evaluation into two objectives:\vspace{0.2em}
\\
\begin{figure*}[!htbp]
\begin{minipage}{0.415\textwidth}
{\footnotesize
    [a] Quantitative Benchmark of Incident Search\\
    \begin{tabular}{|p{0.85cm}|p{0.6cm}|p{0.45cm}|p{0.45cm}|p{0.45cm}|p{0.45cm}|p{0.45cm}|p{0.4cm}|}\hline
    \textbf{Overlap} & \textbf{Metric} & \multicolumn{2}{|c|}{\textbf{Symbolic}} & \multicolumn{2}{|c|}{\textbf{Neural}} & \multicolumn{2}{|c|}{\textbf{Neural+CKG}} \\ \hline
    \multicolumn{2}{|c|}{\textbf{Over Top-10}} & \textbf{Avg} & \textbf{Max} & \textbf{Avg} & \textbf{Max} & \textbf{Avg} & \textbf{Max} \\ \hline
    Document & BLEU & 52.1 & 71.01 & \textbf{54.69} & \textbf{73.71} & \multicolumn{2}{|c|}{\multirow{6}{0.85cm}{Same as Neural}} \\ \cline{1-6}
    Subject & BLEU & 24.79 & 51.59 & \textbf{32.31} & \textbf{58.34} & \multicolumn{2}{|c|}{\ditto} \\ \cline{1-6}
    \multirow{3}{1cm}{Topics} & F1 & 4.13 & 14.0 & \textbf{6.79} & \textbf{16.7}  & \multicolumn{2}{|c|}{} \\ \cline{1-6}
    & BOW & 19.34 & 34.84 & \textbf{21.82} & \textbf{37.48}  & \multicolumn{2}{|c|}{ } \\ \cline{1-6}
    & BLEU & 30.2 & 47.61 & \textbf{32.65} & \textbf{49.91} & \multicolumn{2}{|c|}{\ditto } \\ \cline{1-6}
    Summary & BLEU & 32.65 & 52.24 & \textbf{35.15} & \textbf{54.75}  & \multicolumn{2}{|c|}{} \\ \hline
    \multirow{2}{1cm}{Root Cause} 
    & BOW & 6.43 & 21.41 & 6.87 & 22.21  & \textbf{22.34} & \textbf{35.21}\\ \hline
    & BLEU & 9.69 & 28.52 & 10.31 & 29.11  & \textbf{17.67} & \textbf{45.14}\\ \hline
    \multirow{2}{1cm}{Resolut-ion} 
    & BOW & 5.82 & 23.23 & 6.87 & 24.2  & \textbf{21.42} & \textbf{31.8} \\ \cline{2-8}
    & BLEU & 9.17 & 29.38 & 10.06 & 29.63 & \textbf{16.52} & \textbf{41.27}\\ \hline
    \end{tabular}
    \label{tab:prb_search}}
\end{minipage}
\begin{minipage}{0.287\textwidth}
\begin{tabular}{@{}c@{}}
    \includegraphics[width=\linewidth]{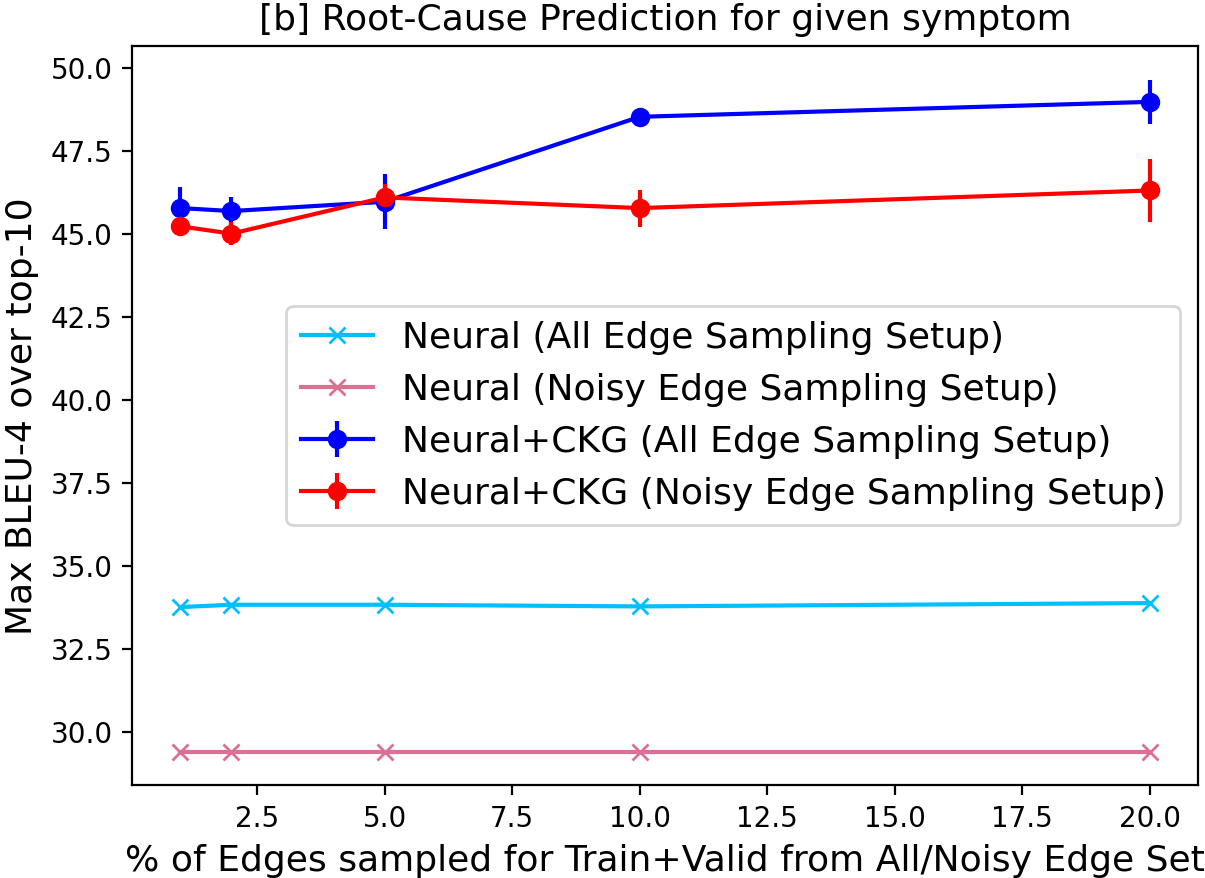}
  \end{tabular}
\end{minipage}
\begin{minipage}{0.287\textwidth}
\begin{tabular}{@{}c@{}}
    \includegraphics[width=\linewidth]{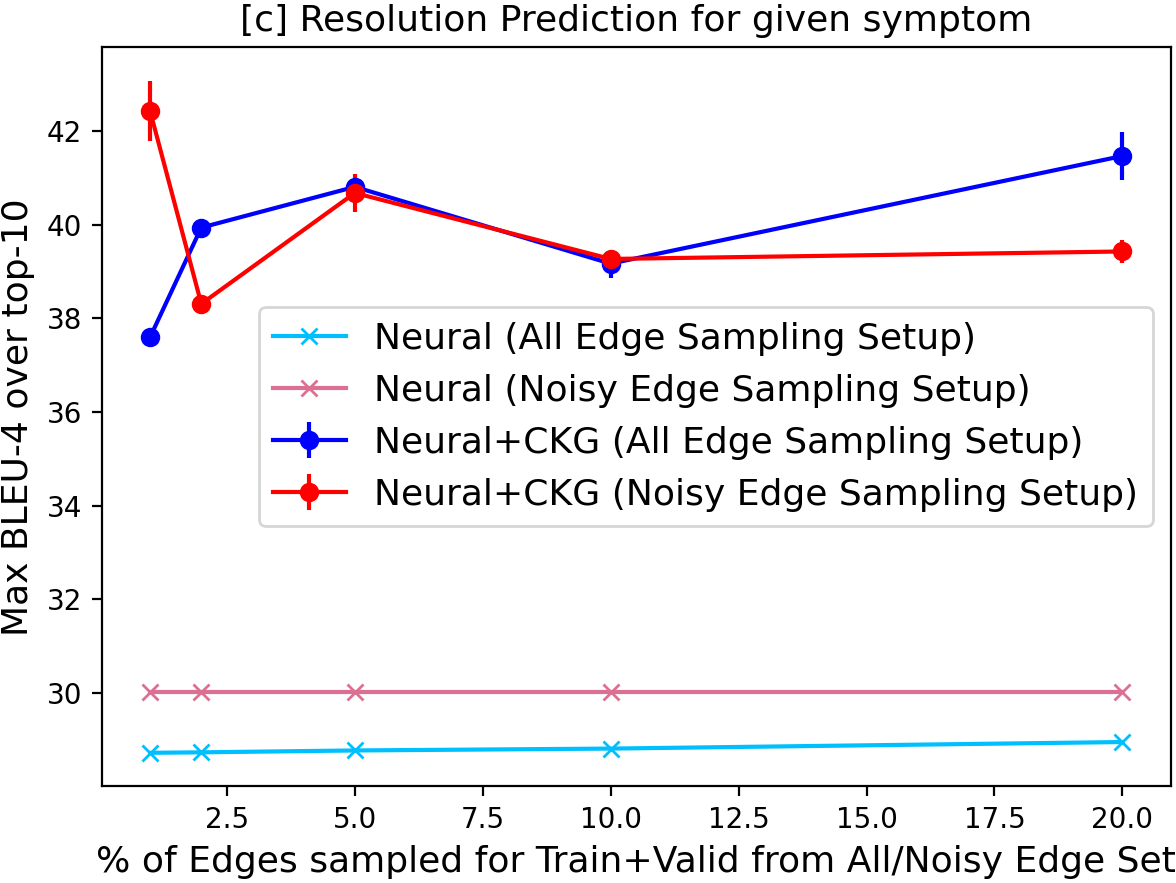}
  \end{tabular}
\end{minipage}\vspace{-1em}
\caption{(a) Quantitative Benchmarking of Incident Search over 2000 PRB records (b) Root Cause (c) Resolution Prediction performance, given the symptom, using Incident Search (IS) and the Causal Knowledge Graph (CKG)}
\label{tab:prb_search}
\end{figure*}
\textbf{i) Evaluating Search:} We compare the retrieved top-k results with the target, by considering either i) the raw PRB Investigation \emph{Document} ii) raw PRB \emph{Subject} iii) extracted \emph{Topics} iv) extracted \emph{Summary}. For each of these we consider BLEU-4gram  \citep{papineni-etal-2002-bleu} similarity, which we compute between a target list (of sentences or phrases) ($D_{t} = [\ldots ,s_i, \ldots]$) and retrieved list ($D_{r} = [\ldots ,s_j, \ldots]$) as $\frac{1}{|D_{t}|}\displaystyle\sum_{s_{i} \in D_{t}} \displaystyle\max_{s_{j} \in D_{r}} BLEU(s_i, s_j)$. Additionally for extracted topics, we also compute i) F1-Score of exact match i.e. matching entire text of target with retrieved and ii) BOW i.e. recall of \textbf{B}ag-\textbf{O}f-NonStop-\textbf{W}ords of the target text over the retrieved. \vspace{0.5em}
\\
\textbf{ii) Evaluating Retrieval based RCA:} We compare the distribution of the top-k likely Root-causes and Resolutions obtained from ICA (as described in Sec \ref{subsec:search}) with the true \emph{Root-Cause} and \emph{Resolution} extracted from the target PRB. For both, we compute the metrics BLEU-4gram, and BOW i.e. recall of \textbf{B}ag-\textbf{O}f-NonStop-\textbf{W}ords. 
\\
For the above metrics, results in Fig. \ref{tab:prb_search}a) are obtained by taking the metric's average or maximum value over the Top-10 results. 
\vspace{0.5em}
\\
\textbf{Models}: We consider different variants of the Incident Search: 
\textbf{i) Neural:} This is the Incident Search as described in Sec \ref{subsec:search}.
\textbf{ii) Symbolic:} This is same as Incident Search, except for replacing the ``neural'' with traditional symbolic search based on the popular user-friendly benchmark Pyserini. For this we index the raw PRB Subject, Investigation, Root Cause and Resolution documents. While exact-match based symbolic search acts as a very strong baseline for us, it should be noted that the search technique itself (neural or symbolic) is not our contribution. Rather our goal is to build a robust pipeline that can handle the noisy non-standardized unstructured documentations created by this agile RCA process.
\textbf{iii) Neural + Causal Knowledge Graph (CKG):} As described in Sec \ref{subsec:search}, this first executes the Incident Search and then runs inference on the Causal Knowledge Graph to further refine the root-cause and resolution prediction. Hence for this model's evaluation would differ from that of Neural Incident Search only in terms of  root-cause and resolution prediction performance. We now briefly elaborate the details of the RCA using the Causal Knowledge Graph.\vspace{0.5em}
\\
\textbf{Using Causal Knowledge Graph in RCA:} Here we train different models for Causal Knowledge Graph based Root Cause and Resolution prediction task, by varying the percentage of symptom-rootcause and symptom-resolution edges in Training data. We create Train+Validation data for these different models by sampling $x$\% (=\{1\%, 2\%, 5\%, 10\%, 20\%\}) of edges from either i) set of All Edges, or ii) set of Noisy edges alone. The first case results in sampling $x$\% edges from True and x\% from Noisy Edges.
These sampled edges are split 70\%:30\% for Train and Validation respectively and the Test data for each of these setups are the set of \emph{remaining} True edges linking symptom-rootcause and symptom-resolution. Based on this we create a separate Test set of PRBs for each setup, consisting of those documents whose symptom-rootcause pair \emph{not} appearing in Train and Validation. Note that for Noisy edge sampling setup, the Test set is same regardless of $x$ (\% of edges for Train+Validation) and is the entire set of 2000 PRB records. Each of these models are then used in an end-to-end style Retrieval based RCA following the technique in \ref{subsec:search}. In Fig \ref{tab:prb_search} a) since the evaluation is done over the entire repository of 2000 PRB records, we use the Noisy edge setup with $x$=1\% i.e. using the least data for training. Further, in Fig \ref{tab:prb_search} b) and c) we respectively compare the root-cause and resolution prediction performance of the Neural and Neural+CKG based Search methods, by varying $x$. For simplicity, we only plot the maximum BLEU-4 score between the top-10 predictions. Since the evaluation set is different for All edge sampling setup, we run both Incident Search (Neural and Neural+CKG) on the same set of PRB records. 
\subsubsection{Our Observations on the results in Fig \ref{tab:prb_search}}:\\
\textbf{Meaningful evaluation metrics:} One striking observation is that the evaluating w.r.t the raw PRB documents achieves high BLEU score throughout. Even with a naive Random Search baseline (which randomly samples top-k  documents) we achieve a high Avg (over top-10) BLEU of 41.0 over PRB document and a meagre 3.94 over PRB Subject. This shows that evaluation over raw PRB investigation documents is meaningless as most of them often contain identical or near-identical template of wordings and common generic details, distracting the search and resulting in high document similarity. Rather evaluating w.r.t the targeted information extracted from them is more accurate reflection of the true model performance.\vspace{0.4em}
\\
\textbf{Symbolic Search:} performs somewhat worser than Neural Search according to these precision-based metrics, on average being 1-2 points weaker in some metrics and upto 3-7 points in others. We also manually observed that Symbolic Search misses some relevant documents simply because of lexical variations like non-canonicalized mentions or agglutinations or typological errors, which are captured by Neural Search. 
So we separately evaluate Neural Search by removing from its retrieved list, any PRB document that has also appeared in the Symbolic Search results. The new top-10 results from the residual set achieve a precision almost as high as that of the original top-10, incurring an average (over all metrics) drop of only 0.93\% and 1.6\% respectively in terms of Avg and Max @top-10.\vspace{0.4em} 
\\
\textbf{Retrieval based RCA}: From Fig \ref{tab:prb_search} b) and c), we observe that both for Root Cause and Resolution prediction, the Incident Search + Causal Knowledge Graph based model far supercedes the Incident Search alone, clearly establishing the efficacy of the Causal Knowledge Graph. One of the main reasons, behind this is that the explicit cluster graph helps impose a structure over the related or similar incidents and learn collectively from them. \vspace{0.4em}
\\
\textbf{RCA Performance Trend by varying $\mathbf{x}$}: For Noisy Edge Sampling setup, since the Test PRB Set is the set of all PRB documents, the results with Incident Search is constant throughout and directly corresponds with the results in Fig \ref{tab:prb_search} a). For All Edge Sampling setup, the Test PRB Set is different for each $x$, but the results with Neural Search remains nearly the same throughout by varying $x$, for both root cause and resolution. Again, in both those cases, Incident Search with Causal Knowledge Graph (CKG) shows overall a mostly positive, but not drastic, trend in performance with increasing Training data size. Even when trained with only 1\% of only noisy edges alone, using CKG gives a significant boost of 15\% (in Max BLEU) in Root Cause and 11\% in Resolution Prediction. \vspace{0.4em}
\\
\textbf{Only Meaningful for Repeated Incidents}: Both for Incident Search and Retrieval based RCA, only repeated incidents will contributing to the precision metrics. The performance is hence limited by the fact that around only 6\% of incidents have atleast 50\% word overlap in terms of Symptom, Root Cause and Resolution (from Fig. \ref{fig:repeat_incidents}) and 4\% incidents are almost identical repeats. 
\subsection{Human Validation \& Case Studies}
\begin{figure*}[!htbp]
\centering 
\includegraphics[width=\textwidth]{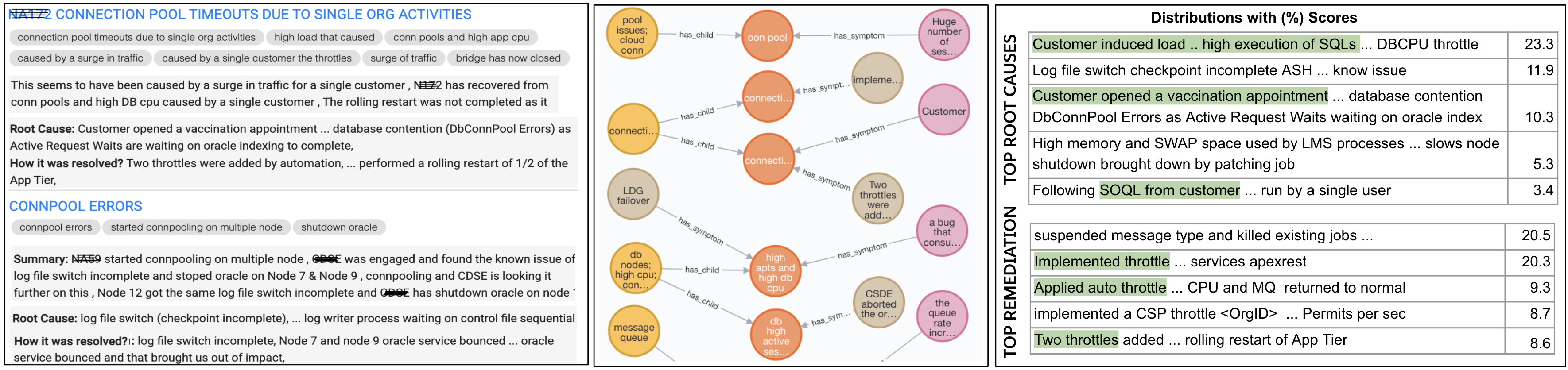}
\caption{ICA output for Incident with Symptom ``Connpool issues'' (i) Incident Search Results (ii) Query Specific Causal Knowledge Subgraph iii) Distribution of top detected root causes and resolutions (Highlighted spans show match with true)}
\label{fig:e2e_example}
\end{figure*}
\begin{table*}[!htbp]
    \centering
    \scalebox{0.9}{
    \begin{tabular}{|p{3.5cm}|p{5.45cm}|p{9cm}|p{0.5cm}|}\hline
    \textbf{Incident Symptom}	& \textbf{True Root Cause (SRE detected)}  & \textbf{ICA predicted Root Cause (Closest Match with True)}	& \textbf{Rank} \\ \hline
custom domain issue &	missing CNAME in validating dns process	& CNAME changes to route the custom origin domain through Edge API	& 3 \\ \hline
single db multi customer 	& high logical IO concurrent singleton Inserts	& diagnostic run aggravated IO issues causing severe IO waits on host	& 2 \\ \hline
search index delayed  &	2 of 3 brokers were down and rebooted  & mq mgmt broker call timeout issue, resolved by bounced mq broker 1,2	& 3 \\ \hline
searchdelayedindex error 	& Bad host hardware caused slow indexing 	& CPU Vendor Hardware.. possible hardware issues.. RAIDCARD failures	& 1 \\ \hline
connpool degradation	& long running aura requests	& Flooded Aura Requests & 	4\\ \hline
high latency active requests	& Service disruption due to db tier	& service disruption due to Process gap in PSU patching .. DB Vendor bug &	4\\ \hline
high latency	& concurent apex errors	& large performance test ...Latency shooted due to concurrent apex errors& 	3\\ \hline
latency	degraded on multiple pods & network switch crosslink failure	& high latency event in app-to-db network traffic due to network inbound limit issue... network packet processing pinned & 	4\\ \hline
high latency high pod cpu &	bulkapi requests from a single user	& customer driven high volume of large malformed concurrent requests & 3\\ \hline
latency degraded in Pod	& caused by disk io on single host	& potential disk iops limitation ... breaching IOPs due to a micro-bursting  & 4\\ \hline
high latency, jetty threads 	& due to index rebuild 	 & An index started being rebuilt which caused the impact &	1\\ \hline 
    \end{tabular}}
    \caption{Real Incident Analysis with Symptom (Input to ICA), True \& ICA Predicted Root Cause and its rank in top-10}
    \label{tab:rca_analysis}
\end{table*}
We now present a more formal evaluation manually validated by domain experts and elaborate on an end-to-end case study below.
\noindent\textbf{Human Validation of Incident Search:} We validated the Incident Search results for 50 handcrafted queries, by domain experts. 32\% of these queries had 1 clause, 56\% had 2 clauses (e.g., \emph{high request rate and high jetty threads}) and 12\% had 3 clauses (e.g. \emph{degraded capacity and connection pools on Single App Server}). We observed that all top-10 results matched atleast 1 query clause, and 53\% and 40\% results respectively matched 2 and 3 query clauses.\vspace{0.3em}
\\
\textbf{Case Study for Real Incident RCA} Our Incident Causation Analysis (ICA) pipeline was recently deployed in production for a pilot study with our Site-Reliability Engineers (SRE) to use in RCA investigation. In this section we take 50 real incidents which occurred post-deployment and do a comparative analysis of the root cause predicted by the ICA pipeline with the true root cause detected by the SRE through their manual investigation. On automatic detection of any anomalous incident, different hand-crafted workflows get auto-triggered to detect the incident symptom. For example, for the incident in Fig \ref{fig:e2e_example} the generated symptom ``\emph{Connpool issues}" is provided as input query to ICA which yields the following outputs for SRE to observe (as in Fig \ref{fig:e2e_example}) during live investigation: i) Incident Search Results, shown in the summarized structured form for quick perusal ii) Query specific Causal Knowledge subgraph to intuitively explore the cluster of incidents or related root causes iii) Retrieval based RCA results which gives the distribution of the top likely root causes and resolutions based on past similar investigations. For e.g. in this incident the true root cause found by SRE was \emph{Performance tests being run by a vaccine cloud customer} which indeed closely corroborated with the top predicted root cause i.e. \emph{Customer opened vaccine appointment causing DB contention}. Consequently the top recommended resolutions are to throttle the URI which typically mitigates this type of issue. 

We performed similar manual comparative analysis on 50 incidents with varying kinds of symptoms. For each incident, the SRE compared the true root cause detected through their manual investigation with the top-K root causes predicted by ICA and judged whether any of the predictions ``match'' the true. Through this analysis, we found that in \textbf{28 out of 50 instances, the root cause predicted by the Retrieval based RCA pipeline of ICA closely matches the true root cause, with the mean rank of the closest matching root-cause being 3.07}. This corroborates quite well with our quantitative results and shows that one of the top-3 results is usually very relevant to the investigation. In Table \ref{tab:rca_analysis} we provide the symptom, and true and predicted root cause of 11 of those 28 incidents. The SRE judged that the remaining 22 incidents were indeed novel with previously unknown root causes. Together our quantitative and qualitative analysis showcases a promising direction in guiding SREs in their RCA investigation based on the knowledge acquired from past investigations. In future this can help in speeding up the investigation by enabling SRE to promptly investigate the root-causes suggested by ICA and infer if an incident is similar to the past incidents and quickly resolve it.

\vspace{0.5em}
\noindent\emph{\textbf{Usability to Cloud Service Incident Management}}
Our ICA pipeline has a minimalistic assumption into the nature of the PRB documents. While incident investigation data can be documented somewhat differently in other cloud industries, their essence remains the same - with details on incident symptom, investigation, root-cause and resolution. For each we assume a completely open-ended form, from which our ICA can extract and re-use relevant information using linguistic cues and only open-source unsupervised and pretrained models (with default hyperparameter settings) the  making it generic enough to be applied to Cloud service Incident Investigations data. 

\vspace{0.5em}
\noindent\textbf{\emph{Threats to Validity}}
Being in a completely unsupervised setting, for evaluating this work, we mostly relied on qualitative evaluation on subset of data and quantitative benchmarking using proxy methods (e.g. incident symptom as query) and analyzing 50 real incidents. We acknowledge that these evaluations cannot exhaustively explore the nature of complications that may arise in real incidents. 
Also since historical knowledge can potentially suffer from knowledge drift and become obsolete. While in our model we handle this by simply notifying users both when any search result or predicted root-cause/resolution are from a substantially old incident, this is a potential future direction open for exploration.

\vspace{-1em}

\section{Conclusion}
\label{sec:conclusion}
In this work, we investigated research on an important and widely available data source in Cloud Service Incident Management, i.e., past incident RCA investigation records (PRB), documented by domain experts. 
But owing to their long unstructured nature of natural language documentation, they have not been much leveraged in AIOps pipelines for RCA yet. In this work, we present ICA, an Incident Causation Analysis engine built at Salesforce, to extract meaningful RCA information from PRB documents, in order to reuse this knowledge for new incidents through Incident Search and retrieval-based RCA. We keep our solution generic by making minimal assumptions about PRB documents and only employing pretrained NLP models in ICA. By doing extensive evaluation and analysis of ICA on PRBs collected in-house over a few years, we showcase that incident reports are indeed a rich goldmine for RCA and our ICA engine is in fact quite effective for improving incident management at least for repeated incidents. Though our current ICA engine serves as a simple but effective retrieval based RCA pipeline, the framework and our models hold promise in future for building a more holistic multimodal multi-source RCA engine.

\bibliographystyle{ACM-Reference-Format}
\bibliography{sample-sigconf}

\end{document}